\documentclass[envcountsect]{llncs}
\usepackage{makeidx}  
\usepackage{amsmath}
\usepackage[pdftex]{graphicx}
\usepackage{url}

\newcommand{\ie}{{\em i.e.}}
\newcommand{\eg}{{\em e.g.}}
\newcommand{\etal}{{\em et al.}}

\newcommand{\comment}[1]{}

\newcommand{\codesize}[1]{\scriptsize #1}
\newcommand{\smallcodesize}[1]{\scriptsize #1}
\newcommand{\code}[1]{{\texttt{\codesize  #1}}}

\newcommand{\subt}{\ensuremath{<:}}

\newcommand{\Kclass}{\texttt{class}}
\newcommand{\Kinterface}{\texttt{interface}}
\newcommand{\Kextends}{\texttt{extends}}
\newcommand{\Kimplements}{\texttt{implements}}
\newcommand{\Kpublic}{\texttt{public}}

\newcommand{\Ksc}{\texttt{;}}
\newcommand{\Kreturn}{\texttt{return}}

\newcommand{\Kland}{\texttt{\&\&}}
\newcommand{\Keqeq}{\texttt{==}}

\newcommand{\env}[1]{\ensuremath{\left\lbrack #1 \right\rbrack}}
\newcommand{\bind}[2]{\ensuremath{#1 / #2}}
\newcommand{\tparam}[1]{\texttt{<}#1\texttt{>}}
\newcommand{\vparam}[1]{\texttt{(}#1\texttt{)}}
\newcommand{\bigoh}[1]{\ensuremath{\mathcal{O}(#1)}}

\newtheorem{thm}{Theorem}[section]
\newtheorem{prop}[thm]{Proposition}
\newtheorem{defn}[thm]{Definition}

\begin{document}
%
\pagestyle{headings}  

\title{Design Pattern-Based Extension of Class Hierarchies to Support Runtime Invariant Checks}

\author{John Lasseter \and John Cipriano}
\institute{Fairfield University, Fairfield CT 06824, USA\\
\email{jlasseter@fairfield.edu},
\email{johnmikecip@gmail.com}}

\maketitle

\begin{abstract}
We present a technique for automatically weaving structural invariant checks into an 
existing collection of classes. Using variations on existing design patterns, we use a 
concise specification to generate from this collection a new set of classes that implement 
the interfaces of the originals, but with the addition of user-specified class invariant 
checks. Our work is notable in the scarcity of assumptions made. Unlike previous design 
pattern approaches to this problem, our technique requires no modification of the original 
source code, relies only on single inheritance, and does not require that the attributes 
used in the checks be publicly visible. We are able to instrument a wide variety of class 
hierarchies, including those with pure interfaces, abstract classes and classes with type 
parameters. We have implemented the construction as an Eclipse plug-in for Java 
development.

\end{abstract}

%
%

\section{Introduction}

Several, if not most, mainstream languages include features to support object-oriented programming, yet most of 
these (C++, C\#, Java, Python, etc.) lack any native language support for the specification and runtime checking 
of class invariants.  While it is usually easy enough to implement the invariant predicates themselves, manual 
addition imposes further requirements in order to implement the operational requirements of invariant checking 
and to handle the interplay of invariant specification and inheritance.  Class invariants are further troublesome 
in that they involve direct access to an object's attributes. This makes manual addition particularly 
unappealing, as the available choices are invasive with respect to the original interface and implementation (to 
which we may not have access), compromise encapsulation, and are error-prone if done manually.  

This paper presents a lightweight, non-invasive technique for automatically extending a collection of class 
definitions with a corresponding collection of structural invariant checks.  The invariants are given as a stand-
alone specification, which is woven together with the original source files to produce a new collection of drop-
in replacement classes that are behaviorally indistinguishable from the originals in the absence of invariant-
related faults but will expose such faults in a way that the original classes do not.  Each replacement is 
defined to be a subclass (indeed, a {\em subtype} \cite{liskov94toplas}) of the original class whose 
functionality it extends, and it can thus be substituted in any context in which the original occurs.  
%
The generation is itself completely automatic, and the incorporation into a test harness or other program is 
nearly seamless.   We focus here on the Java language, a choice that complicates the overall strategy in some 
ways while simplifying it in others.

\section{Background and Related Work}

A {\em class invariant} is a conjunction of predicates defined on the values of an object's individual 
attributes and on the relationships between them.  It characterizes an object's ``legal'' states, giving the 
predicates that must hold if the object is to represent an instance of that abstraction.  Usually, a class 
invariant is given in conjunction with the {\em contracts} for each publicly-visible method of a class, \ie, the 
preconditions that must hold on arguments to each method call and the consequent guarantees that are made as 
postconditions upon the method's return.   Unlike the contracts, however, a class invariant is a property 
concerning only an object's {\em data values}, even (especially) when those values are not publicly visible.  An 
invariant must hold at every point between the object's observable actions, \ie\ upon creation of any object 
that is an instance of this class and both before and after every publicly-visible method call \cite
{liskov01pdj,meyer97oosc}.  At other points, including non-visible method calls, it need not hold, and runtime 
checks are  disabled in this case.   Further, since runtime invariant checks can impose a non-trivial 
performance penalty on a system, in general, it is desirable to have a mechanism for leaving the checks in place 
during testing, while removing them from a final, production system.  Finally, there is an important interplay 
between the subtype relation (which determines when one object can safely be substituted in a context calling 
for another \cite{liskov94toplas}) and class invariants:  if $B$ is a subtype of $A$ (as well as a subclass) 
then the invariant for $B$ must include all of the constraints in $A$'s invariant 
\cite{liskov01pdj,meyer97oosc}.

Some languages offer native support for invariant checking, but for  Java and other languages that lack this, 
including such checks is challenging.  A common approach is to make use of the language's assertion mechanism, 
by  including assertions of the invariant  at the end of each constructor body and at the beginning and end of the 
body of each public method \cite{liskov01pdj}. If the language's assertions mechanism is used, disabling the 
checking functionality after testing is usually quite easy.  However, this approach carries the disadvantage of 
requiring the class designer to code not only the predicates themselves but also an explicit handling of the 
inheritance requirements and the full execution model, discussed above.  Both of these tasks must be implemented 
for each invariant definition, in each class.

To avoid the implementation burden of the assertions approach, we can use a tool that generates the invariant 
checks from either specialized annotations of the source code \cite{flanagan02pldi,leavens06sen,liu03sea} or 
reserved method signatures \cite{karaorman98techrep,prasetya07techrep,prasetya08issta}.   Essentially, such tools 
offer language extensions to resemble native support for invariant definitions.  In comparison to assertion-based 
approaches, they eliminate the requirement of implementing the execution model, a clear advantage.  As with the 
assertions approach, annotation approaches are invasive, in that they require modification of the original source 
code.  More substantially, the approach generally requires the use of a specialized, nonstandard compiler, whose 
development may not keep up with that of the language%
\footnote{For example, JML has not seen active development since version 1.4 of the Java language \cite{jml12website}.}.

Instead, we can view the addition of runtime invariant checking across a class {\em hierarchy} as a kind of 
cross-cutting concern, \ie{} code that is defined across several classes and hence resists encapsulation.  Under 
this view, it is natural to approach this problem as one of aspect-oriented programming (AOP) \cite
{kiczales97ecoop}, in which we can use a tool such as AspectJ \cite{AspectJGuide} to define the checks separately 
as aspects.  The entry and exit points of each method become the join points, the point cuts are inferred from a 
class's method signatures, and the invariant check itself becomes the advice \cite
{briand05icsm,skotiniotis04oopslac}.     Unlike annotation-based approaches, aspect weaving can be done without 
the need for a non-standard compiler, either through source code transformation or byte code instrumentation 
\cite{czarnecki00gpBook}.  However, the AOP approach also presents several difficulties.  For example, Balzer 
\etal{}  note that mainstream tools such as AspectJ lack a mechanism to enforce the requirement  that the  
definition of a class's invariant include the invariant of its parent class \cite{balzer05rise}.  It is possible 
to write  invariant checking ``advice'' so  that it correctly calls the parent class's invariant check, but this 
must be done manually (\eg{} \cite{briand05icsm}).  A similar problem occurs in implementing  the correct 
disabling of checks on non-public calls.  Lastly, because aspects cannot in general be prevented from changing an 
object's state, the weaving of additional aspects may compose poorly with the aspect that provides the invariant 
check \cite{agostinho08splat,balzer05rise,klaeren00gcse}. It is possible that another aspect could  break the 
class invariant, and since interleaving of multiple aspects is difficult to control, it is possible the two 
aspects could interleave in such a way as to make the invariant failure go undetected.

The work closest in spirit to our own is the design pattern approach of Gibbs, Malloy, and Power (\cite
{gibbs02ase,malloy06stvr}.   Targeting development in the C++ language, they present a choice of  two patterns 
for weaving a separate specification of invariant checks into a class hierarchy, based on the well known {\em 
decorator} and {\em visitor} patterns \cite{gamma95designpatterns}.  However, the decorator approach involves a 
fairly substantial refactoring of the original source code.  Moreover, the authors note that this technique 
interacts poorly with the need to structure invariant checks across a full class hierarchy.  The refactoring in 
this case is complex, and it requires the use of multiple inheritance to relate the decorated classes 
appropriately, making it unsuitable for languages such as Java, which support only single inheritance.  Their 
alternative is an application of the visitor pattern, in which the invariant checks are implemented as the {\em 
visit} methods in a single {\em Visitor} class.  This pattern usually requires that the classes on the ``data 
side'' implement an {\em accept} method, which is used to dispatch the appropriate {\em visit} method, but in 
their use of it, only the top of the class hierarchy is modified to be a subclass of an ``invariant 
facilitator'', which handles all {\em accept} implementations.  However, successful implementation of the {\em 
visit} methods rests on the assumption that all fields are either publicly visible or have their values readily 
available through the existence of accessor (``getter'') methods.   Unless the language simply lacks a mechanism 
to hide this representation (\eg{} Python), such exposure is unlikely to be the case, as it violates 
encapsulation, permitting uncontrolled manipulation of an object's parts, either directly or through aliasing 
\cite{liskov01pdj}. 

The central thesis of our work is that, under assumptions common to Java and other statically-typed OO languages, 
these limitations---source code modification, multiple inheritance, and public accessibility of fields---are 
unnecessary for a design-pattern approach.  The remainder of the present paper shows how to relax them.

\section{Weaving Invariant Checking from Specifications}\label{main}

Our approach draws from the Gibbs/Malloy/Power design pattern efforts and from ideas in AOP in the treatment of 
invariant specifications as a cross-cutting concern.  We begin with an assumption that the class invariants are 
given in a single specification file, separate from classes that they document.  Each constraint is a boolean-
valued Java expression, with the invariant taken to be the conjunction of these expressions.  We assume (though 
do not hope to enforce) that these expressions are free of side effects, and that the invariant given for a child 
class does not contradict any predicates in inherited invariants.   Otherwise, the particulars of the 
specification format are unimportant.  The current version of our tool uses  JSON \cite{json12website}, but any 
format for semi-structured data will do.

We focus on the Java programming language, which means that we assume a statically-typed, object-oriented 
language, with introspective reflection capabilities, support for type parameters in class definitions, single 
inheritance (though implementation of multiple interfaces is possible), and a uniform model of virtual method 
dispatch.  We make some simplifications of the full problem.  Specifically, we work only with synchronization-
free, single-threaded, non-{\em final} class definitions, and  we consider only instance methods of a class that 
admit overriding, \ie{}, non-{\em static}, non-{\em final}%
\footnote{The {\em final} keyword has two uses in Java:  to declare single-assignment, read-only variables and to 
prohibit extension of classes or overriding of methods.  The latter form is equivalent to the {\em sealed} 
keyword in C\#, and it is this usage we avoid here. }
 method definitions.
We do not consider anonymous inner class constructs nor the {\em lambda expressions} planned for Java 8 \cite
{java12jsr335}.  Finally, we assume a class's field visibility grants at least access through inheritance (\ie{} 
{\em protected} accessbility or higher).  This last is made purely for the sake of simplifying the technical 
presentation, since, as discussed in section \ref{concl}, introspection makes it easy to handle variables of any accessibility.

\subsection{An Inheritance-Based Approach}\label{invcheck:inh-approach}

As a first effort, we will try an approach that leverages the mechanism of inheritance and the redefinition of 
inherited method signatures through subtyping polymorphism.  The idea is to derive from a class and its invariant 
a subclass, in which we wrap the invariant in a new, non-public method (perhaps with additional error reporting 
features), similar to the ``{\em repOK}'' approach advocated by Liskov and Guttag \cite{liskov01pdj}.  To this 
new subclass, we also add methods $\phi_1$ and $\phi_2$ to handle the checking tasks at (respectively) method 
entry and exit points, and we use these to define constructors and overridden versions of every public method.

Let $A$ be a class, with parametric type expression $T_A$ defined on type parameters $S_A$,
field declarations $\overrightarrow{\tau ~a}$, invariant $\rho_A$, constructor definition 
$A\vparam{\overrightarrow{\tau_A ~y}}$ and public method $\tau_f~ f\vparam{\overrightarrow{\sigma_f ~z}}$.  

\medskip
{\codesize
\begin{tabular}{l}
\Kpublic ~\Kclass ~$A$\tparam{$T_A$} \{\vspace{3pt}\\ 
\quad $\overrightarrow{\tau ~a;}$\vspace{3pt}\\ 
\quad \Kpublic ~$A$\vparam{$\overrightarrow{\tau_A ~x}$}~\{ ~ \ldots ~ \}\vspace{3pt}\\ 
\quad \Kpublic ~$\tau_{f_A}~ f_A$\vparam{$\overrightarrow{\sigma_{f_A} ~y}$}~\{ ~ \ldots ~ \} \vspace{3pt}\\ 
\}\\
\end{tabular}
}
\bigskip

\noindent We extend $A$ with runtime checking of $\rho_A$ by generating the subclass in Fig. 
\ref{fig:naive-inh}, where $T_{A'}$ and ${S_{A'}}$ are identical to $T_A$ and 
${S_A}$ (respectively), except perhaps for renaming of type parameters (\ie, they are $\alpha$-equivalent).

\begin{figure}[t]
\begin{center}
{\smallcodesize
\begin{tabular}{l}
\code{\Kpublic ~\Kclass ~$A'$\tparam{$T_{A'}$} ~\Kextends~ $A$\tparam{${S_{A'}}$}
                                                                                           \{}\vspace{3pt}\\ 
\quad            \code{private int $\delta$ = 0;} \vspace{3pt}\\
\quad            \code{public ~$A$\vparam{$\overrightarrow{\tau_A ~x}$}~\{} \\
\quad\quad            \code{super\vparam{$\overrightarrow{x}$};}\\
\quad\quad            \code{$\delta$ = $\delta$ + 1;}\\
\quad\quad            \code{$\phi_2$();}\\
\quad            \code{\}}\vspace{3pt}\\ 
\quad            \code{public ~$\tau_{f_A}~ {f_A}$\vparam{$\overrightarrow{\sigma_{f_A} ~y}$}~\{ }\\
\quad\quad            \code{$\phi_1$();} 
~~                    \code{$\tau_{f_A}$ $\chi$ =  super.${f_A}\vparam{\overrightarrow{y}}$;} 
~~                    \code{$\phi_2$();}\\
\quad\quad            \code{return $\chi$;}\\
\quad            \code{\}} \vspace{3pt}\\ 
\quad            \code{private boolean $inv$() \{ return~$\rho_A$;  \}} \vspace{3pt}\\ 
\quad            \code{private void $\phi_1$() \{}\\
\quad\quad            \code{if ($\delta$ == 0 \Kland{} !$inv$())} \\
\quad\quad\quad           {\em $\langle$ handle invariant failure $\rangle$ }\\
\quad\quad            \code{$\delta$ = $\delta$ + 1;} \\
\quad            \code{\}} \vspace{3pt}\\ 
\quad            \code{private void $\phi_2$() \{}\\
\quad\quad             \code{$\delta$ = $\delta$ - 1;} \\
\quad\quad             \code{if ($\delta$ == 0 \Kland{} !$inv$())} \\
\quad\quad\quad           {\em $\langle$ handle invariant failure $\rangle$ }\\
\quad            \code{\}} \vspace{3pt}\\ 
\code{\}}\\
\end{tabular}
}
\caption{Inheritance-based generation of invariant checks}
\label{fig:naive-inh}
\end{center} 
\end{figure}

For each constructor in $A'$, the body executes the ``real'' statements of the corresponding superclass 
constructor, followed by a check of $\rho_A$, whose execution is itself controlled by the $\phi_2$ method.  
Likewise, the body of each public method $f_A$ wraps a call to the superclass's version between checks of 
$\rho_A$, with execution controlled by the $\phi_1$ and $\phi_2$ methods.  If $f_A$ returns a value, then this 
value is captured in the overridden version in a ``result'' variable, $\chi$.  A method or constructor call is 
publicly-visible precisely when the call stack depth on a given $A'$ object is 0, and this value is tracked by 
the additional integer-valued field $\delta$.  The $\phi_1$ and $\phi_2$ methods increment/decrement $\delta$ as 
appropriate, evaluating $\rho_A$ only if $\delta = 0$.%
\footnote{In the presence of concurrency, we would need a more sophisticated mechanism;  keeping track of the 
call stack depth on an object for each thread, synchronizing all method calls on the object's monitor lock, and 
so on.}

The  inheritance-based approach suggests an easy mechanism for reusing code while adding the necessary 
invariant checks and capturing the distinction between publicly-visible and inner method calls.  For the user, 
the burden consists of replacing constructor calls to $A$ with the corresponding calls for $A'$.   This may be an 
excessive requirement when $A$ objects are used in production-level code, but in many settings where invariant 
checking is desirable, such constructor calls are limited to only a handful of sites.  In the JUnit framework, 
for example, integration of $A'$ objects into unit tests for $A$ is likely quite simple, as object construction 
occurs mainly in the body of a single method, {\em setUp}.

Note the assumptions of uniform polymorphic dispatch and non-{\em final} declarations here.  If a class cannot be 
extended (\eg{} {\em String} and other objects in the {\em java.lang} package), then construction of a subclass 
that implements the invariant checks is obviously impossible.  Similarly, a method whose dispatch is statically 
determined cannot be transparently overridden, and if declared {\em final}, it cannot be overridden at all.  In 
many languages (notably, C\# and C++) the default convention is {\em static} dispatch, with dynamic binding 
requiring an explicit {\em virtual} designation;  in such cases, the inheritance construction is far less 
convenient and may be impossible without some refactoring of the original source code.  

Unfortunately, our first attempt fails in two critical ways, which becomes apparent when we attempt to construct 
the invariant-checking extension across a  hierarchy of class definitions.  First of all, the inheritance 
hierarchy of a collection of objects requires a corresponding structure in the composition of invariant checks.  
This problem is very similar to the one encountered in the ``decorator'' approach of \cite{malloy06stvr}, but the 
multiple-inheritance solution given there is unavailable in a single-inheritance language such as Java.  Consider 
a class $B$ that is a subtype of $A$ (written $B \subt A$):

\medskip
{\codesize
\begin{tabular}{l}
\Kpublic ~\Kclass ~$B$\tparam{$T_B$} \Kextends ~ $A$\tparam{${S_B}$}\{\vspace{3pt}\\ 
\quad $\overrightarrow{\tau~b}$\vspace{3pt}\\ 
\quad \Kpublic ~$B$\vparam{$\overrightarrow{\tau_B ~y}$}~\{ ~ \ldots ~ \}\vspace{3pt}\\ 
\quad \Kpublic ~$\tau_{g_B}~ g$\vparam{$\overrightarrow{\sigma_{g_B} ~z}$}~\{ ~ \ldots ~ \} \vspace{3pt}\\ 
\}\\
\end{tabular}
}

\begin{figure}[t]
\begin{center}
\includegraphics[scale=0.6]{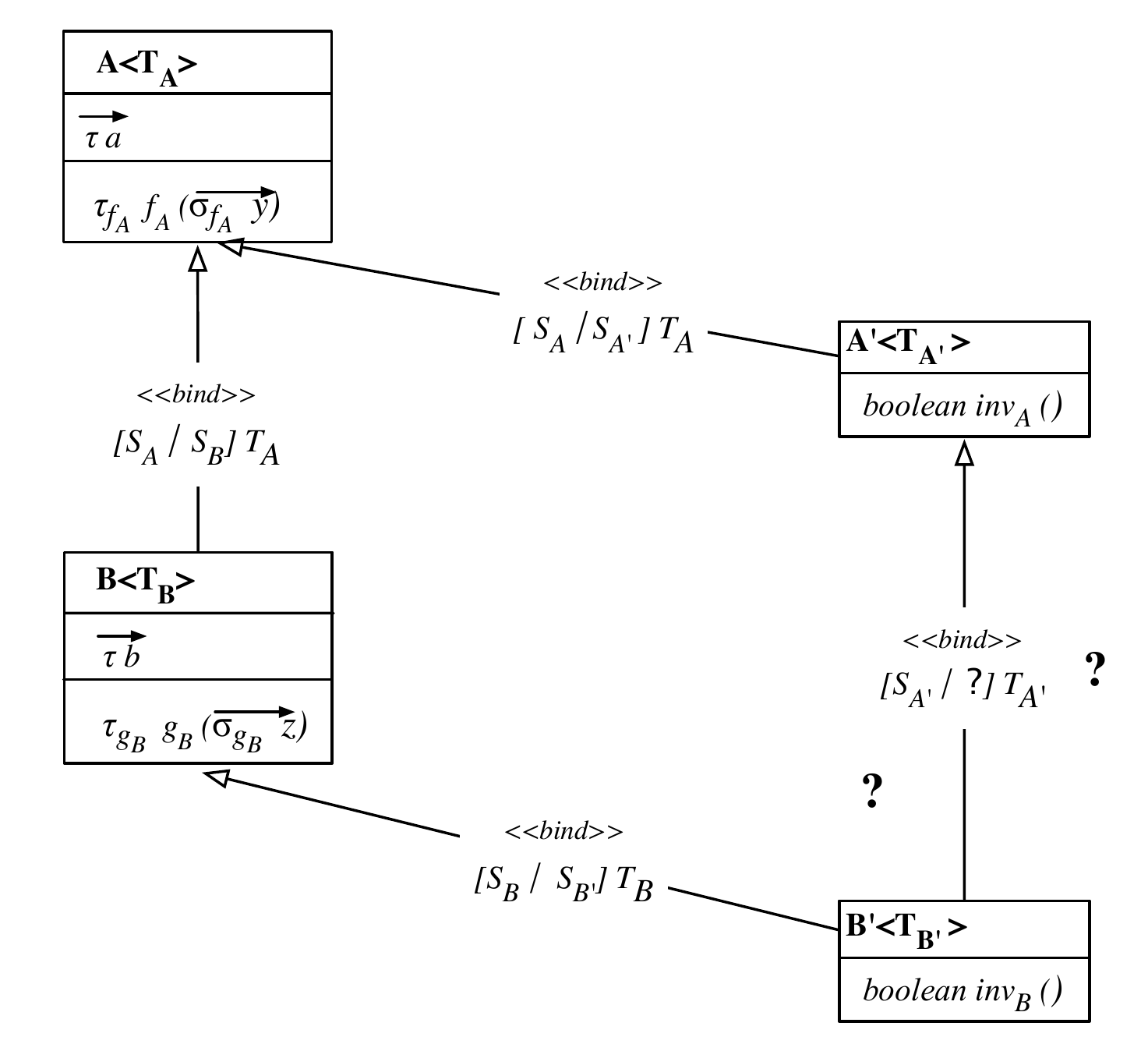}
\caption{Design flaw in the naive inheritance approach}
\label{fig:naive-limit}
\end{center} 
\end{figure}

\noindent Figure \ref{fig:naive-limit} depicts the problem%
\footnote{There and throughout this paper, we write $\env{\bind{S}{\tau}} T$ to denote the substitution of type 
expression $\tau$ for the type parameter $S$ in expression $T$, and use the shorthand 
$\env{\bind{S_1}{\tau_1},\bind{S_2}{\tau_2}} T$  to denote the composition of type expressions 
$\env{\bind{S_1}{\tau_1}}\env{\bind{S_2}{\tau_2}} T$.}.
The invariant for a $B$ object, $inv_B$, must include the $A$ invariant---\ie{}, $inv_B = inv_A \land \rho_B$.   
However, a $B'$ object cannot access the fields of its associated $B$ object through inheritance and also reuse 
the functionality of the $inv_A$ method.  We might choose to have $B'$ descend from $A'$ instead, but this only 
works if all fields in $B$ are publicly accessible.  As discussed above, this is unlikely to be the case.  

The second, related failure is that inheritance does not facilitate a correct binding of the type parameters.  
Again, this is clear from Fig. \ref{fig:naive-limit}.  An instantiation of $B$ supplies a type $\tau$ to the 
parameters ${S_{B}}$, which is used in turn to bind the parameters ${S_{A}}$ with 
argument $\env{\bind{S_B}{\tau},\bind{S_A}{S_B}} T_{A}$.  When we instantiate $B'$ instead, this same $\tau$ 
binds the parameters ${S_{B'}}$, with the resulting chain of arguments binding $A$'s parameters 
${S_A}$ as $\env{\bind{S_{B'}}{\tau}, \bind{S_B}{S_{B'}},\bind{S_A}{S_B}} T_A$.  
For correct use of the $A'$ invariant check in this $B'\langle\tau\rangle$ object, we would need to bind the 
type parameter of $A'$, ${S_{A'}}$, in the same way we do $A$'s parameter, $S_A$;  
\ie{} with argument $\env{\bind{S_{B'}}{\tau}, \bind{S_{A'}}{S_{B'}}, \bind{S_A}{S_{A'}}} T_{A}$,  a binding 
that cannot be ensured, unless $B'$ is a subclass of $A'$.

\subsection{Exposing the Representation}\label{ssec:exposure}

Though unsuccessful on its own, we can use the inheritance approach of Section \ref{invcheck:inh-approach} as the 
basis for an auxiliary pattern, which we call an {\em exposure pattern}.  The idea is to construct from the 
original hierarchy a corresponding set of classes that offers the interface of the original collection and in 
addition, a controlled exposure of each object's representation.  The  machinery for checking the invariants is 
factored into separate classes, as discussed in Section \ref{invcheck}, below.

Consider a class definition 

\bigskip
{\codesize
\begin{tabular}{l}
\Kpublic ~\Kclass ~$A$\tparam{$T_A$} \{\vspace{3pt}\\ 
\quad $\tau_1 ~a_1\Ksc \quad\ldots\quad \tau_k ~a_k\Ksc$\vspace{3pt}\\ 
\quad \Kpublic ~$A$\vparam{$\overrightarrow{\tau_A ~y}$}~\{ ~ \ldots ~ \}\vspace{3pt}\\ 
\quad \Kpublic ~$\tau_{f_A}~ {f_A}\vparam{\overrightarrow{\sigma_{f_A} ~z}}$~\{ ~ \ldots ~ \} \vspace{3pt}\\ 
\}\\
\end{tabular}
}
\bigskip

\noindent 
We derive the {\em exposure} interface

\medskip
{\codesize
\begin{tabular}{l}
\code{\Kpublic ~\Kinterface ~$IA_E$\tparam{$T_{A''}$}  \{}\vspace{3pt}\\ 
\quad            $\tau_1 ~\gamma_{a_1}\code{()}\Ksc$ \vspace{3pt}\\ 
\quad\quad       \ldots \vspace{3pt}\\ 
\quad            $\tau_m ~\gamma_{a_m}\code{()}\Ksc$ \vspace{3pt}\\ 
\code{\}}\\
\end{tabular}
}

\bigskip

\noindent and {\em exposed} class 

\medskip
{\codesize
\begin{tabular}{l}
\code{\Kpublic~\Kclass~$A_E$\tparam{$T_{A'}$}
                     ~\Kextends~ $A$\tparam{${S_{A'}}$}
                     ~\Kimplements~$IA_E$\tparam{${S_{A'}}$}
                                                                                           \{}\vspace{3pt}\\ 
\quad            \code{private int $\delta$ = 0;} \vspace{3pt}\\
\quad            \code{private void $\phi_1$() \{} \ldots  \code{\}} \vspace{3pt}\\ 
\quad            \code{private void $\phi_2$() \{} \ldots  \code{\}} \vspace{3pt}\\ 
\quad            \code{protected boolean $inv$(}$InvV ~v$\code{) \{} \ldots  \code{\}} \vspace{12pt}\\ 
\quad            \code{public ~$A_E$\vparam{$\overrightarrow{\tau_A ~y}$}~\{} \\
\quad\quad            \code{super\vparam{$\overrightarrow{y}$}; ~ $\delta$ = $\delta$ + 1; ~ $\phi_2$();}\\
\quad            \code{\}}\vspace{3pt}\\ 
\quad            \code{public ~$\tau_{f_A}~ {f_A}\vparam{\overrightarrow{\sigma_{f_A} ~y}}$~\{ }\\
\quad\quad            \code{$\phi_1$(); ~ $\tau_{f_A}$ $\chi$ =  super.${f_A}\vparam{\overrightarrow{y}}$; 
                           ~$\phi_2$();}\\
\quad\quad            \code{return $\chi$;}\\
\quad            \code{\}} \vspace{18pt}\\ 
\quad            $\Kpublic~ \tau_1 ~\gamma_{a_1}\code{()} \{ ~ \Kreturn~a_1\Ksc ~ \}$ \vspace{3pt}\\ 
\quad\quad       \ldots \vspace{3pt}\\ 
\quad            $\Kpublic~ \tau_m ~\gamma_{a_m}\code{()} \{ ~ \Kreturn~a_m\Ksc ~ \}$ \vspace{3pt}\\ 
\code{\}}\\
\end{tabular}
}
\bigskip

\noindent where $T_{A'}$, $T_{A''}$ and ${S_{A'}}$, ${S_{A''}}$ are 
$\alpha$-equivalent to 
$T_{A}$ and ${S_{A}}$, as above.  Note that the fields $a_1 \ldots a_m$ include all of the 
original $a_1 \ldots a_k$ and perhaps others, as discussed on page \pageref{prop:type-correctness}, below.  The 
constructors and public methods in $A_E$ are overridden in exactly the same manner as in the $A'$ class of 
Section \ref{invcheck:inh-approach}, and likewise the implementation of the $\phi_1()$ and $\phi_2()$ methods. 
The representation exposure happens through the $\gamma_{a_i}()$, a set of raw ``getter'' methods that expose 
each of the object's fields.  In the presence of inheritance, the corresponding structure is realized not in the 
derived class but in the derived {\em interfaces}.  Thus, for example,

\medskip
{\codesize
\begin{tabular}{l}
\Kpublic ~\Kclass ~$B$\tparam{$T_B$} \Kextends ~ $A$\tparam{${S_B}$}\{\vspace{3pt}\\ 
\quad $\tau_1 ~b_1\Ksc \quad\ldots\quad \tau_l ~b_l\Ksc$\vspace{3pt}\\ 
\quad \Kpublic ~$B$\vparam{$\overrightarrow{\tau_B ~y}$}~\{ ~ \ldots ~ \}\vspace{3pt}\\ 
\quad \Kpublic ~$\tau_{g_B}~ f$\vparam{$\overrightarrow{\sigma_{g_B} ~z}$}~\{ ~ \ldots ~ \} \vspace{3pt}\\ 
\}\\
\end{tabular}
}
\bigskip

\noindent gives rise to the interface and class definitions

\medskip
{\codesize
\begin{tabular}{l}
\code{\Kpublic ~\Kinterface ~$IB_E$\tparam{$T_{B''}$}~ \Kextends 
                                                ~ $IA_E$\tparam{${S_{B''}}$} \{}\vspace{3pt}\\ 
\quad            $\tau_1 ~\gamma_{b_1}\code{()}\Ksc$ \vspace{3pt}\\ 
\quad\quad       \ldots \vspace{3pt}\\ 
\quad            $\tau_n ~\gamma_{b_n}\code{()}\Ksc$ \vspace{3pt}\\ 
\code{\}}\\
\end{tabular}
}

\medskip
{\codesize
\begin{tabular}{l}
\code{\Kpublic~\Kclass~$B_E$\tparam{$T_{B'}$} \Kextends~$B$\tparam{${S_{B'}}$}
                     \Kimplements ~ $IB_E$\tparam{${S_{B'}}$} \{ } \ldots ~ \}

\end{tabular}
}
\bigskip

\begin{figure}[t]
\begin{center}
\includegraphics[scale=0.5]{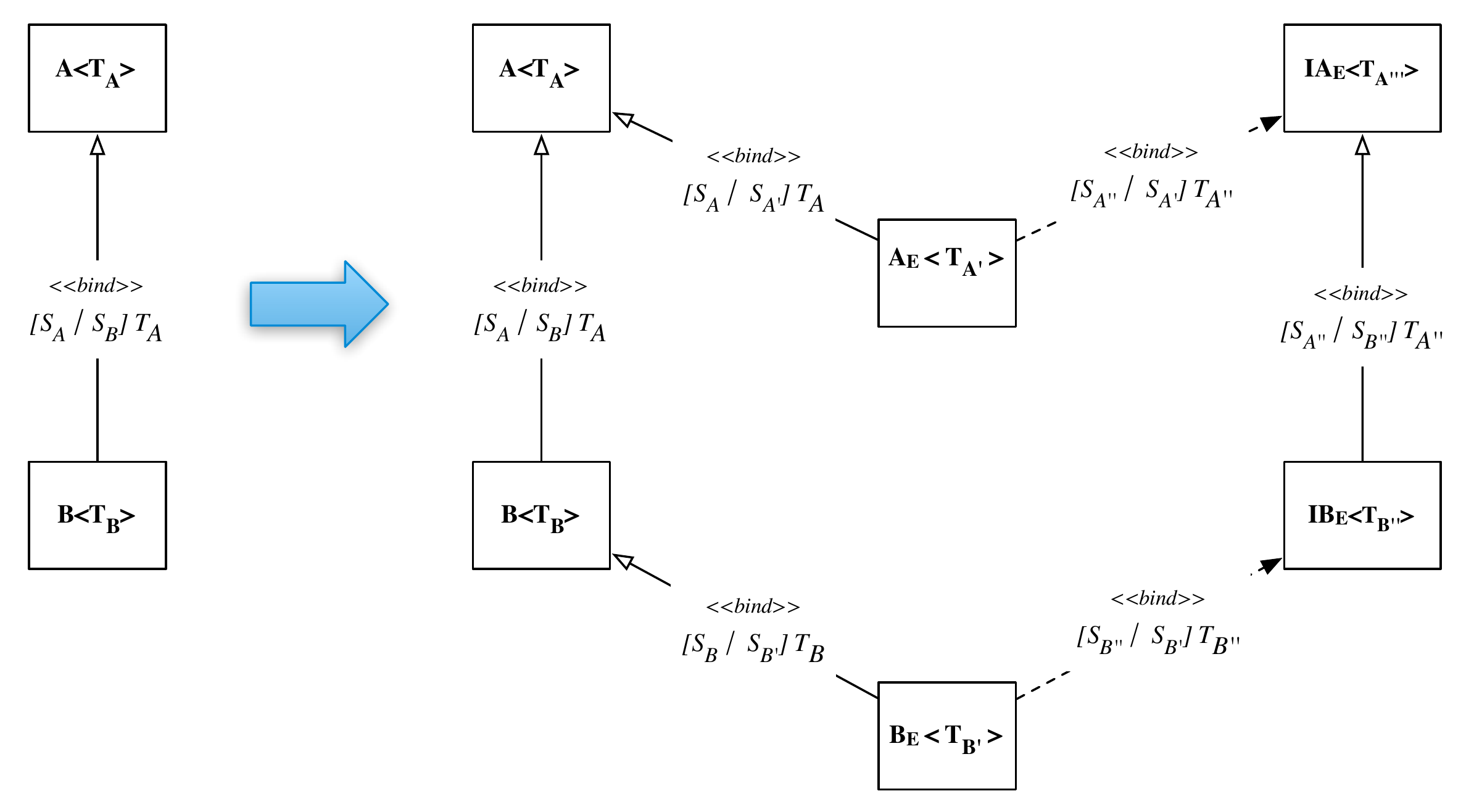}
\caption{Exposure pattern construction}
\label{fig:exposure-pattern}
\end{center} 
\end{figure}

\noindent The construction is illustrated in Fig. \ref{fig:exposure-pattern}.

\subsubsection{Correctness}\label{sssection:correctness}

Since the type expressions in a class definition are copied to its exposed class and interface (perhaps with 
$\alpha$-renaming of the parameters), it is easy to see that

\begin{prop}\label{prop:type-correctness}
For any type expression $\tau$, an instance of a class $A$ has type $A\tparam{T(\tau)}$ if and only if $A_E$ and 
$IA_E$ have types $A_E\tparam{T(\tau)}$ and $IA_E\tparam{T(\tau)}$, respectively.
\qed
\end{prop}

The construction of the accessor methods is less obvious.  While we construct $\gamma_{a_i}()$ for each of the 
fields $\{a_1, \ldots, a_k\}$, we may need to construct others, as well, in case the invariant $\rho_A$ makes 
reference to any inherited fields for which we have not already constructed an interface.  This can happen in the 
case of an incomplete specification of the class hierarchy and invariants.  The simplest way to handle this is to 
include in the interface a $\gamma_{a_i}()$ for each declared field in the corresponding $A$ classes and also for 
each variable that occurs without explicit declaration in the the predicate $\rho_A$.  However, we can leverage 
the inheritance of interfaces to eliminate redundant declarations (though not implementations, as discussed 
below).

To make the construction precise, we denote the {\em free variables} of the predicate $\rho_A$ by $FV(\rho_A)$, 
\ie{} those variables that occur in $\rho_A$ without being explicitly declared in $\rho_A$.  Conversely, the 
{\em bound} variables in a class $A$, $BV(A)$, are the instance fields declared in $A$. %
%
The following definition captures the notion of variables that are ``free'' in $A$ through inheritance:

\newcommand{\inh}[1]{\ensuremath{\mathcal{I}(#1)}}
\begin{defn}\label{defn:inh}
Let $P$ be a specification of a collection of classes and their associated invariants.  For a class $A$, 
{\em the set of fields exposed through inheritance in $A$}, $\inh{A}$, is defined by

\[
\inh{A} = 
         \begin{cases}
         \emptyset &, \textrm{if $A$ has no superclass specified in $P$} \\
         \inh{C} \cup BV(C) \cup FV(\rho_C)&, \textrm{if $A \subt C$ and $C$ is specified in $P$}
         \end{cases}
\]
\end{defn}

\noindent We use this to define the necessary method signatures in each exposure interface.  

\begin{defn}\label{defn:exp-interface}
Given class $A$ and invariant $\rho_A$, the body of $IA_E$ consists of the the signatures
\[
IA_E = \left\{ \tau_{a_i} \gamma_{a_i}() ; ~|~ a_i \in BV(A) \cup FV(\rho_A) \setminus \inh{A} \right\}
\]
\noindent where each $\tau_{a_i}$ is the declared type of $a_i$.
\end{defn}

\begin{defn}\label{defn:succ-exposure}
For a field, $\tau_{a_i}~a_i$, either declared in or inherited by a class $A$, we say that $a_i$ is 
{\em successfully exposed for A} if either

\begin{itemize}
\item there is an interface $IA_E$ and subclass 

{\codesize
\begin{tabular}{l}
\Kclass~$A_E~\Kextends~A~\Kimplements~IA_E$ 
\end{tabular}
}

\noindent such that $IA_E$ includes a method interface

{\codesize
\begin{tabular}{l}
$\tau~\gamma_{a_i}$();
\end{tabular}
}

\noindent and for every $A_E$ object $o$, $o.\gamma_{a_i}() \Keqeq~o.a_i$

\item $A$ is a subclass of $C$, and $a_i$ is successfully exposed for $C$.
\end{itemize}
\end{defn}

Given $A$ and $\rho_A$, the construction for $IA_E$ in Definition \ref{defn:exp-interface} and the accompanying 
implementation $A_E$ combine to give us the representation exposure we need for $\rho_A$.  In particular,

\begin{prop}\label{prop:exposure-correctness}
If $x \in FV(\rho_A)$, then $x$ is successfully exposed for $A$.
\qed
\end{prop}

\subsubsection{Space Requirements}\label{ss:space}

The primary difference between the exposure pattern construction and the inheritance-based effort of Section \ref
{invcheck:inh-approach} lies in the construction of the exposure interfaces, whose inheritance structure is 
congruent to that of the original collection of classes.  Like the earlier attempt, however, the collection of 
exposed {\em classes} does not share this same relation, and as a consequence, both approaches are subject to 
some unfortunate redundancy consequences.  In particular, we cannot reuse code between distinct exposed classes, 
even when the classes they expose are related by inheritance.  For example, if a class $A$ contains fields $a_1$ 
and $a_2$ and public method $f()$ then the exposed class $A_E$ must override $f()$, and it must include exposure 
methods $\gamma_{a_1}$ and $\gamma_{a_2}$, according to the interface $IA_E$.  If $B \subt A$ contains fields 
$b_1$, $b_2$, and method $g()$, then it must override not only $g()$ but also $f()$, with the body of the 
overridden $f()$ identical to that in $A_E$.  Likewise, it must implement not only the $\gamma_{b_1}$ and 
$\gamma_{b_2}$ methods from the $IB_E$ interface, but also $\gamma_{a_1}$ and $\gamma_{a_2}$.  

Happily, all of this is easily automated, and it is reasonable to suppose the space overhead manageable.   Note 
first that, with the exception of classes at the top of a specified hierarchy, the size of the interface 
generated for a class is proportional to the number of fields in that class.  Recalling Definitions 
\ref{defn:inh} and \ref{defn:exp-interface}, we can see that this is so because 

\begin{prop}\label{prop:inh-trivial}
Let $C$ be a class included in a specification $P$.  For every class $A \subt C$, 
$FV(\rho_A) \setminus \inh{A} = BV(A)$.
\end{prop}

In other words, only for classes specified at the top of an inheritance hierarchy will we ever need to generate  
additional $\gamma$ declarations in the corresponding interfaces.  In all other cases, the accessor interfaces  
for inherited fields are inherited from the corresponding parent interfaces.  Hence, the space required to extend 
a collection of classes depends only on the size of each class and the depth of the inheritance relationship in 
the collection.  Specifically, if we assume a bound of $n$ new field and method definitions on each class and an 
inheritance depth of $h$, then the overall space growth is given by 

\[
\sum_{i=1}^{h} \left(\sum_{j=1}^{i} n \right) = \left( \sum_{i=1}^{h} i  \right) n = 
       \left( \frac{h(h+1)}{2} \right) n \in \bigoh{h^2 n}
\]

\noindent It is difficult to give a general characterization of either $n$ or $h$, but there is reason to suspect 
that both are manageable values in practice.   McConnell recommends a limit of 7 new method definitions in a 
class \cite{mcconnell04codecomplete}.  Shatnawi's study \cite{shatnawi10tse} finds no significant threshold value 
for $h$.  Classes in the JDK's {\em java.*} and {\em javax.*} libraries implement anywhere from less than 10 to 
over 100 new methods, while the largest depth of any inheritance tree is 8.  

\subsection{Adding the Invariant Checks}\label{invcheck}

As in Gibbs/Malloy/Power \cite{malloy06stvr}, we implement the runtime invariant checks themselves through an 
application of the {\em visitor pattern} \cite{gamma95designpatterns}, in which the methods implementing the 
invariant checks are aggregated into a single class (the ``visitor''), with the appropriate method called from 
within the class being checked (the ``acceptor'').  Unlike their approach, however, our exposure pattern allows 
us to do this without modification of any part of the original source files, not even at the top of the 
inheritance hierarchy.  

Suppose we have a class $A\tparam{T_A}$, with invariant $\rho_A$.  From these, we generate the exposed class
$A_E\tparam{T_{A'}}$ and the exposure interface $IA_E\tparam{T_{A''}}$, as in Section \ref{ssec:exposure}.  The 
specification of $\rho_A$ and the access methods defined for $IA_E$ are used to generate an invariant checking 
``visitor'' class:

\bigskip
{\codesize
\begin{tabular}{l}
\Kpublic ~\Kclass ~$InvV$ \{\vspace{3pt}\\ 
\quad \Kpublic  ~\tparam{$T_{A_I}$} 
                          ~ \code{void}~ $v_A$\vparam{$IA_E\tparam{{S_{A_I}}}~ obj$}~\{ \\
\quad\quad  $\tau_1~ a_1$ \code{~=~obj.$\gamma_1$\code();} \\
\quad\quad\quad  \ldots \\
\quad\quad   $\tau_n~ a_n$ \code{~=~obj.$\gamma_n$\code();} \vspace{6pt}\\

\quad\quad  $\langle\langle$ {\em compute $\rho_A$ and return the result} $\rangle\rangle$ \\
\quad\} \vspace{3pt}\\ 
\}\\
\end{tabular}
}
\bigskip

\noindent where $T_{A_I}$  and ${S_{A_I}}$ are equivalent to $T_A$ and its parameters 
${S_{A}}$, as above.

Runtime checking of $\rho_A$ is invoked in the $A_E$ methods through calls to that class's $inv$ method, which 
serves as the ``accept'' method, handling dispatch of the appropriate invariant check:

\bigskip
{\codesize
\begin{tabular}{l}
\code{\Kpublic~\Kclass~$A_E$\tparam{${S_{A'}}$}~\Kextends~ $A$\tparam{${S_{A'}}$}
                     ~\Kimplements~$IA_E$\tparam{${S_{A'}}$}
                                                                                           \{}\vspace{3pt}\\ 
\quad            \code{private int $\delta$ = 0;} \vspace{3pt}\\
\quad            \code{private void $\phi_1$() \{} \ldots \code{\}} \vspace{3pt}\\ 
\quad            \code{private void $\phi_2$() \{}  \ldots \code{\}} \vspace{3pt}
                 \code{~~//} (as defined in Section \ref{invcheck:inh-approach})
\\ 
\quad            \code{private boolean $inv$(}$InvV ~v$\code{) \{} \\
\quad\quad             $v.v_A$\code{(this);}\\
\quad\quad             \Kreturn~$v$\code{.valid();}\\
\quad  \code{\}} \vspace{6pt}\\ 
\quad\quad  \ldots \vspace{6pt}\\
\code{\}}\\
\end{tabular}
}
\bigskip

\noindent Note that each $v_A()$ method in $InvV$ takes an argument of type $IA_E$ and not $A_E$.  This is 
necessary, because of the need to compose an invariant check with that of the object's superclass in each 
invariant method.  For example, if we have $B \subt A$, we define $v_B()$ as

\medskip
{\codesize
\begin{tabular}{l}
\quad \Kpublic  ~\tparam{$T_B$} 
                          ~\code{void}~ $v_B$\vparam{$IB_E\tparam{{S_B}}~ obj$}~\{ \vspace{3pt}\\
\quad\quad          $v_A$\code{( (}$IA_E\tparam{{S_B}}$\code{ ) obj);} \vspace{3pt}\\
\quad\quad          $\langle\langle$ {\em compute $\rho_B$, as above} $\rangle\rangle$ \vspace{3pt}\\
\quad  \code{\}} 
\end{tabular}
}
\bigskip
        
\noindent Since $A_E$ and $B_E$ are not related by inheritance, it would not be possible to directly cast 
\code{obj} to its superclass's exposed version.  Fortunately, the interface is all we need.  

Finally, although we structure our solution here according to the traditional visitor pattern 
conventions, we do not really need the full generality of that pattern.  In particular, it is unnecessary to 
support full double dispatch, as we only need one instance of $InvV$, and no $v_i()$ method will ever invoke a 
call back to the $inv()$ method of an object (not even indirectly, since the $\phi_1$ and $\phi_2$ methods in a 
class prevent a call to $inv()$ if one is already running).  Our implementation of this approach as an Eclipse 
plugin instead drops the $InvV$ parameter from every $inv$ method, relying instead on a single, static instance 
of the invariant visitor:

\bigskip
{\codesize
\begin{tabular}{l}
\quad            \code{private boolean $inv$() \{} \\
\quad\quad             $InvV v$\code{~=~}$InvV$\code{.getInstance();} \\
\quad\quad             \ldots \\
\quad  \code{\}} \vspace{6pt}\\ 
\end{tabular}
}

\section{Example:  Unit Testing}\label{example}

Method contracts and class invariants are particularly useful in testing.  In combination with test oracles, 
the use of runtime invariant and pre/post-conditions checks improves the exposure of faults as well as the 
diagnosability of faults when they are detected \cite{briand03spe,letraon06tse}.  Our implementation as an 
Eclipse plug has proven useful in diagnosing invariant-related faults.

For example, a simple {\tt List} interface provides an abstraction for the list data type.  A standard way to 
implement this is with an underlying doubly-linked list, in which we keep a pair of ``sentinel'' head and tail 
nodes, with the ``real'' nodes in the list linked in between:

{\codesize
\begin{verbatim}
    public abstract class AbstractList<T> implements List<T> {
        protected int size;
        ...
    }
    
    public class DLinkedList<T> extends AbstractList<T> implements List<T> {
        // inherited from AbstractList:  int size
        protected DNode<T> head,  tail;  ...
    }
\end{verbatim}
}

\noindent Among other predicates, the invariant for {\em DLinkedList} requires that $\forall n \neq tail$, 
$n.next.prev = n$.

This was given as part of a project for the first author's data structures course, and among the student 
submissions received was this implementation of {\em remove()}, in which the {\em cur.prev} pointer is not 
correctly updated:

{\codesize
\begin{verbatim}
    public boolean remove(T v) {
        DNode<T> cur = head.next;        
        while (cur != tail) {
            if (cur.data.equals(v)) {
                DNode<T> prev = cur.prev;  cur = cur.next;  prev.next = cur;
                size--;                
                return true;
            } else 
               cur = cur.next;
        }           
        return false;
    }
\end{verbatim}
}

A JUnit test suite failed to uncover this fault, passing this and the tests for 12 other methods:

{\codesize
\begin{verbatim}
    public void testRemove() {
        ls.add("a"); ls.add("b"); ls.add("c"); ls.add("d"); ls.add("a"); ls.add("d"); 
        int sz = ls.size();
        assertTrue(ls.remove("a"));    assertTrue(ls.size() == sz - 1);
        sz = ls.size();        
        assertTrue(!ls.remove("**"));    assertTrue(ls.size() == sz);
    }
\end{verbatim}
}

From the original source code and a specification of invariants our tool generates the classes and interfaces 

{\codesize
\begin{verbatim}
    public interface IExposedAbstractList<T> {
        int _getSize();
    }
    public interface IExposedDLinkedList<T> extends IExposedAbstractList<T> {
        DNode<T> _getHead();
        DNode<T> _getTail();
    }
    
    public abstract class ExposedAbstractList<T>
                           extends AbstractList<T> implements IExposedAbstractList<T> { ... }  
                  
    public class ExposedDLinkedList<T> 
                           extends DLinkedList<T> implements IExposedDLinkedList<T> { ... }
            
    public class RepOKVisitor {
        ...
        public <T> void visit(IExposedAbstractList<T> _inst) { ... }
        public <T> void visit(IExposedDLinkedList<T> _inst) { ... }
        ...
    }
\end{verbatim}
}

\noindent Objects in a JUnit test suite are constructed in the {\em setUp()} method, and 
a simple modification was all that was needed to cause {\em testRemove()} to fail appropriately:

{\codesize
\begin{verbatim}
    protected void setUp() {
    //    ls = new DLinkedList<String>();
        ls = new ExposedDLinkedList<String>();
    }
\end{verbatim}
}

\section{Conclusion and Future Work}\label{concl}

The design pattern given here provides a fairly seamless approach for adding correct runtime invariant checking 
to a class hierarchy, through the construction of drop-in replacements that can be removed as easily as inserted.      
In addition to the core material presented here, there are a number of extensions possible.  

For example, the presentation in this paper relies on the assumption above that all fields in a class are 
accessible through inheritance.  Happily, this is an easy if tedious limitation to overcome.  If instead the 
field is declared with only intra-object or intra-class access (\eg{} Java's ``{\tt private}''), we can use the 
introspective capabilities of the language to manufacture a locally-visible {\em get} method.  To access a 
\code{private}  field $x$, for example, our implementation generates a $\gamma_{x}$ that handles the unwieldy 
details of Java introspection:

\bigskip
{\codesize
\begin{tabular}{l}

                \code{private $\tau$~$\_getX$() \{}   \\
\quad                 \code{Class klass = this.getClass();}   
\quad
                      \code{Field field = null;}   \\
\quad                 \code{while (field == null) \{}   \\
\quad\quad                 \code{try \{}   \\
\quad\quad\quad                 \code{field = klass.getDeclaredField("$x$");}   
          \quad                 \code{field.setAccessible(true);}   \\
\quad\quad                 \code{\} catch (NoSuchFieldException e) \{}   \\
\quad\quad\quad                 \code{klass = klass.getSuperclass();}   \\
\quad\quad                 \code{\}}   \\
\quad                 \code{\}}   \\
\quad                 \code{$\tau$~$x$ = null;}   \\
\quad                 \code{try \{}    \\
\quad\quad                 \code{$x$ = ($\tau$) field.get(this);}   \\
\quad                 \code{\} catch (IllegalAccessException e) \{}   
     \quad                 \code{e.printStackTrace();}   
     \quad                 \code{throw new Error();}   
\quad                 \code{\}}    \\
\quad                 \code{return $x$;}   \\
                \code{\}}   \\
\end{tabular}
}
\bigskip

Other extensions, such as the 
inclusion of anonymous inner classes, concurrency, or {\em final} classes/methods, remain as open challenges.

Finally, the work described here incorporates only the invariant checks, rather than full contracts, and it would 
clearly be useful to extend our design pattern to support this.  While we conjecture that our technique is easily 
extendable to this purpose, the invariant checks present the most interesting problems, owing to their need for 
attribute access and hierarchical definition.  Philosophically, ordinary unit testing already performs at least 
the behavioral components of contract checking, \ie{} the checks of pre and post-conditions.  What unit testing 
cannot do is determine whether the invariant continues to hold, as it is often impossible to access an object's 
fields.  The difference lies in the fact that both pre and post conditions are inherently extensional 
specifications.  They impose requirements on method arguments and return values, but on the object itself, all 
constraints are made upon the abstraction of the object, not the concrete implementation.  That implementation---
whose consistency with the abstraction is the core assertion of a class invariant---is 
by definition opaque to an object's user.

\subsubsection{Acknowledgments}
The ideas in this paper began with an assignment in the first author's Spring 2010 Software Construction class, 
and the students there provided valuable feedback. Our thanks also to Prof. Peter Boothe of Manhattan College, 
for help in analyzing the inheritance and method complexity of the JDK.


\bibliographystyle{splncs}

\end{document}